\documentstyle{article}

\input pz.sty
\input epsf.sty

\begin{document}

\PZhead{1}{29}{2009}{22 October}

\PZtitletl{Photometric observations and modeling}{of type IIb supernova 
2008\lowercase{ax}}

\PZauth{D.Yu. Tsvetkov$^1$, I.M. Volkov$^{1,2}$, P.V. Baklanov$^3$,
S.I. Blinnikov$^{3,1}$, O. Tuchin$^4$} 
\PZinst{Sternberg Astronomical Institute, University Ave. 13,
119992 Moscow, Russia}
\PZinst{Astronomical Institute of the Slovak Academy of Sciences, 059 60 
Tatranska Lomnica, Slovak Republic}
\PZinst{Institute for Theoretical and Experimental Physics, 117259, Moscow, 
Russia}
\PZinst{Samara Palace of Children's Creativity, Kujbyshev str. 151,
443010 Samara, Russia}

\begin{abstract}

CCD {\it UBVRI} photometry is presented for type IIb SN 2008ax for about
320 days. The photometric behavior 
is typical for core-collapse SNe with low amount of hydrogen.
The main photometric parameters are derived and the 
comparison with SNe of similar types is reported. Preliminary modeling 
is carried out, and the results are compared to the observed light 
curves. The main parameters of the hydrodynamical models are close to those
used for SN IIb 1993J. 

\end{abstract}

\medskip
\medskip

\PZsubtitle{Introduction}

Supernova SN 2008ax was discovered independently by
Mostardi et al. (2008) and Itagaki (Nakano and Itagaki, 2008) on March
3.45 UT and March 4.62 UT, respectively.
The magnitude of SN at discovery, estimated on unfiltered
CCD frames, was about 16. 
The first detection was only 
6 hours after the image with limiting magnitude about 18.5 
and showing no sign of the SN was obtained by Arbour (2008).
The offsets from the nucleus of the host galaxy NGC 4490 are
53$\arcs$.1E, 25$\arcs$.8S. 
The projected distance from the center is 2.8 kpc, while the
radius of the galaxy is about 9 kpc.
NGC 4490 is a barred spiral galaxy of type SBcd.
SN II-P 1982F was discovered earlier in this galaxy
(Tsvetkov, 1984).
The positions of two SNe are quite close, SN 1982F occurred 19$\arcs$
(0.9 kpc) closer to the nucleus at about the same positional angle 
(116$\deg$ for SN 2008ax, 120$\deg$ for SN 1982F). 

Crockett et al. (2008) identified a source coincident with the position
of SN 2008ax in pre-explosion HST observations in three optical filters.
The possible progenitor may be a single massive star (initial mass
$\sim 28 {\rm M}_{\odot}$), which loses most of its H-rich envelope
and explodes as an 11-12 M$_{\odot}$ helium-rich Wolf-Rayet star,
or an interacting binary producing a stripped progenitor.

Photometric and spectroscopic observations of SN 2008ax covering
first 2 months past discovery were reported by Pastorello et al. (2008)
(hereafter P08). The object displayed typical spectral and photometric
evolution of a type IIb supernova, consistent with the explosion
of a young Wolf-Rayet star. 

Roming et al. (2009) (hereafter R09) presented UV, optical, X-ray and radio 
properties
of SN 2008ax.
They detected initial fading in UV light curves followed by a rise,
reminiscent to the dip seen in type IIb SN 1993J.

\newpage
\PZsubtitle{Observations and reductions}

We started photometric monitoring of SN 2008ax 4 days after the
discovery and continued observations until 2009 January 23. 
CCD images in {\it UBVRI} filters were obtained with the following 
instruments: the 50-cm reflector of Astronomical Institute
of Slovak Academy of Sciences at Tatranska Lomnica with SBIG ST-10XME
CCD camera (hereafter S50); the 50-cm meniscus telescope and
the 60-cm reflector of Crimean Observatory of Sternberg Astronomical Institute
equipped respectively with Meade Pictor 416XT and Apogee AP-47 cameras
(C50, C60); the 70-cm reflector of Sternberg Astronomical Institute
in Moscow with Apogee AP-7 CCD camera (M70); the 1-m reflector of Simeiz
Observatory with AP-47 camera (C100). The images on 2009 January 23 were
obtained at the 2-m Faulkes Telescope North (F200).

The standard image reductions and photometry were made using IRAF.\PZfm 
\PZfoot{IRAF is distributed by the National Optical Astronomy Observatory,
which is operated by AURA under cooperative agreement with the
National Science Foundation}

The galaxy background around SN 2008ax is strong and non-uniform, 
and we applied image subtraction for most of the frames.  
Our observations did not allow to construct good template frames, and
we used for subtraction the images of NGC 4490 downloaded from
the ING archive.\PZfm\PZfoot{http://casu.ast.cam.ac.uk/
casuadc/archives/ingarch}

After subtraction, the magnitudes of the SN
were derived by PSF-fitting relative to 
a sequence of local standard stars. The comparison stars are shown on
Fig. 1, and their magnitudes are reported in Table 1.
\PZfig{11cm}{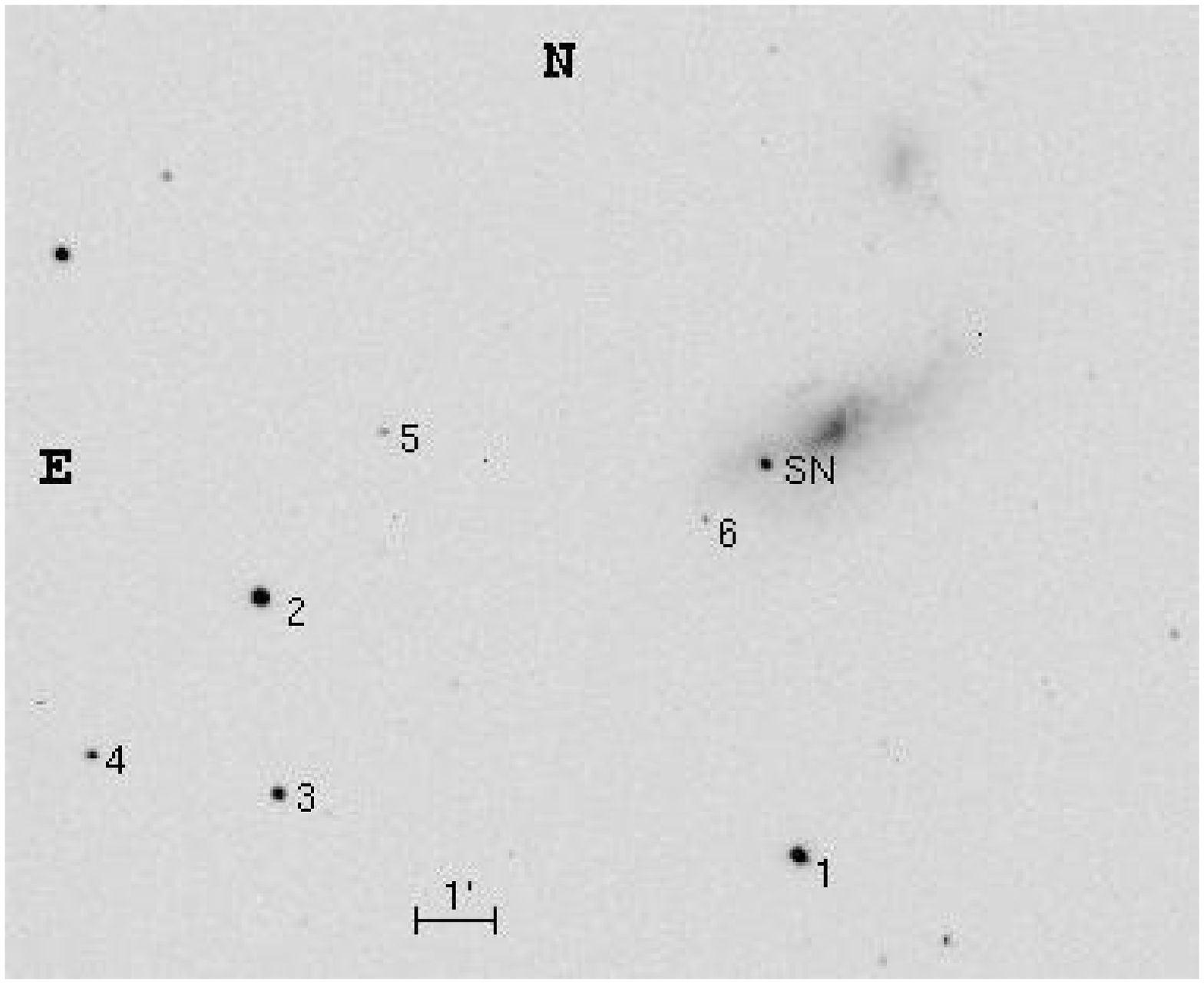}{SN 2008ax with local standard
stars}
\begin{table}
\centering
\caption{Magnitudes of local standard stars}\vskip2mm
\begin{tabular}{ccccccccccc}
\hline
Star & $U$ & $\sigma_U$& $B$ & $\sigma_B$ & $V$ & $\sigma_V$ 
& $R$ & $\sigma_R$ & $I$ & $\sigma_I$
\\
\hline
1 &   11.73 &0.04&  11.62& 0.01 & 11.06& 0.01 & 10.73& 0.01 & 10.41& 0.01\\
2 &   13.78 &0.05&  12.94& 0.01 & 11.89& 0.01 & 11.35& 0.01 & 10.85& 0.01\\
3 &   13.66 &0.05&  13.71& 0.01 & 13.18& 0.01 & 12.83& 0.02 & 12.50& 0.02\\
4 &   14.61 &0.06&  14.60& 0.02 & 13.96& 0.01 & 13.55& 0.04 & 13.18& 0.02\\
5 &         &    &  16.25& 0.03 & 15.20& 0.02 & 14.57& 0.03 & 14.03& 0.03\\
6 &   16.58 &0.07&  16.02& 0.03 & 15.00& 0.02 & 14.47& 0.02 & 13.93& 0.03\\
\hline
\end{tabular}
\end{table}

Stars 1-4 were measured photoelectrically in the $B,V$ filters by
Tsvetkov (1984), the comparison with the new CCD data reveals good
agreement, the mean differences are $\Delta B = -0.02; \Delta V = 0.04$ mag.
We may conclude that both calibrations are sufficiently correct. 

The results of observations of the SN are presented in Table 2.

\begin{table}
\caption{Observations of SN 2008ax}\vskip2mm
\begin{tabular}{cccccccccccl}
\hline
JD 2454000+ & $U$ & $\sigma_U$ & 
$B$ & $\sigma_B$ & $V$ & $\sigma_V$ & $R$ & $\sigma_R$ &
$I$ & $\sigma_I$ & Tel.\\
\hline
532.51&      &     & 17.25& 0.08&  15.92& 0.03 & 15.35& 0.04&  15.09& 0.04& S50 \\
537.59&      &     & 15.22& 0.02&  14.54& 0.02 & 14.19& 0.03&  13.84& 0.04& S50 \\
541.37& 14.43& 0.06& 14.46& 0.03&  13.88& 0.03 & 13.67& 0.05&  13.37& 0.03& S50 \\
544.42& 14.09& 0.06& 14.22& 0.02&  13.68& 0.02 & 13.36& 0.03&  13.06& 0.02& S50 \\
546.39& 14.06& 0.06& 14.13& 0.02&  13.56& 0.02 & 13.24& 0.04&  12.93& 0.02& S50 \\
551.31& 14.33& 0.06& 14.19& 0.02&  13.41& 0.02 & 13.08& 0.03&  12.73& 0.03& S50 \\
552.53& 14.56& 0.05& 14.25& 0.02&  13.45& 0.01 & 13.07& 0.02&  12.73& 0.02& S50 \\
553.43& 15.07& 0.12& 14.38& 0.05&  13.58& 0.04 & 13.10& 0.07&  12.80& 0.03& S50 \\
555.52& 15.21& 0.08& 14.67& 0.06&  13.64& 0.05 & 13.12& 0.06&  12.86& 0.05& S50 \\
556.42& 15.60& 0.08& 14.82& 0.05&  13.77& 0.05 & 13.32& 0.05&  12.87& 0.05& S50 \\
557.41& 15.87& 0.07& 15.02& 0.02&  13.87& 0.02 & 13.27& 0.05&  12.90& 0.03& S50 \\
563.40& 17.27& 0.13& 15.77& 0.04&  14.39& 0.03 & 13.76& 0.05&  13.19& 0.04& S50 \\
564.47&      &     & 15.81& 0.07&  14.41& 0.03 & 13.75& 0.03&  13.18& 0.03& S50 \\
570.37&      &     & 16.21& 0.10&  14.71& 0.02 & 14.05& 0.05&  13.40& 0.03& S50 \\
579.38&      &     & 16.45& 0.04&  14.97& 0.02 & 14.33& 0.02&  13.68& 0.03& M70 \\
583.33&      &     & 16.52& 0.05&  15.07& 0.03 & 14.46& 0.03&  13.79& 0.04& M70 \\
585.35&      &     & 16.49& 0.05&  15.05& 0.03 & 14.42& 0.03&  13.77& 0.04& M70 \\
590.30&      &     & 16.62& 0.05&  15.20& 0.03 & 14.59& 0.03&  13.89& 0.03& M70 \\
601.47&      &     & 16.84& 0.09&  15.41& 0.04 & 14.86& 0.04&  14.08& 0.03& S50 \\
602.34&      &     & 16.99& 0.04&  15.50& 0.03 & 14.89& 0.02&  14.17& 0.03& M70 \\
613.32&      &     & 16.91& 0.04&  15.65& 0.02 & 15.08& 0.02&  14.35& 0.03& M70 \\
616.41&      &     & 17.03& 0.04&  15.69& 0.03 & 15.08& 0.04&  14.35& 0.04& S50 \\
623.33&      &     & 16.96& 0.05&  15.87& 0.03 & 15.31& 0.02&  14.56& 0.03& M70 \\
628.44&      &     & 17.25& 0.10&  16.00& 0.03 & 15.45& 0.03&  14.60& 0.04& S50 \\
643.34&      &     &      &     &  16.37& 0.03 & 15.73& 0.03&  14.97& 0.06& C50 \\
644.32&      &     &      &     &  16.35& 0.03 & 15.75& 0.03&  15.07& 0.11& C50 \\
647.31&      &     &      &     &  16.35& 0.04 & 15.80& 0.03&       &     & C50 \\
647.31&      &     & 17.53& 0.05&  16.41& 0.02 & 15.85& 0.02&  15.00& 0.04& C100\\
649.30&      &     &      &     &  16.41& 0.03 & 15.83& 0.03&       &     & C50 \\
656.30&      &     &      &     &  16.63& 0.02 & 15.98& 0.03&       &     & C50 \\
658.30&      &     &      &     &  16.63& 0.10 & 16.00& 0.05&       &     & C50 \\
660.33&      &     &      &     &  16.67& 0.04 & 16.07& 0.03&  15.49& 0.10& C60 \\
674.34&      &     & 18.25& 0.14&  17.01& 0.06 & 16.33& 0.04&  15.55& 0.03& S50 \\
675.32&      &     &      &     &  17.01& 0.04 & 16.45& 0.07&  15.73& 0.08& S50 \\
677.32&      &     &      &     &  17.19& 0.05 & 16.36& 0.05&  15.60& 0.05& S50 \\
679.32&      &     &      &     &  17.31& 0.07 & 16.52& 0.06&  15.73& 0.16& S50 \\
699.28&      &     &      &     &  17.57& 0.20 & 16.78& 0.07&       &     & M70 \\
720.24&      &     &      &     &       &      & 16.99& 0.07&       &     & C60 \\
781.63&      &     &      &     &       &      & 17.90& 0.05&       &     & C60 \\
783.62&      &     &      &     &  18.98& 0.07 & 18.18& 0.06&       &     & C60 \\
855.05&      &     & 20.90& 0.18&  20.45& 0.07 & 19.22& 0.03&  18.77& 0.03& F200 \\
\hline
\end{tabular}
\end{table}

\bigskip
\bigskip
\medskip

\PZsubtitle{Light and color curves}

\PZfig{13cm}{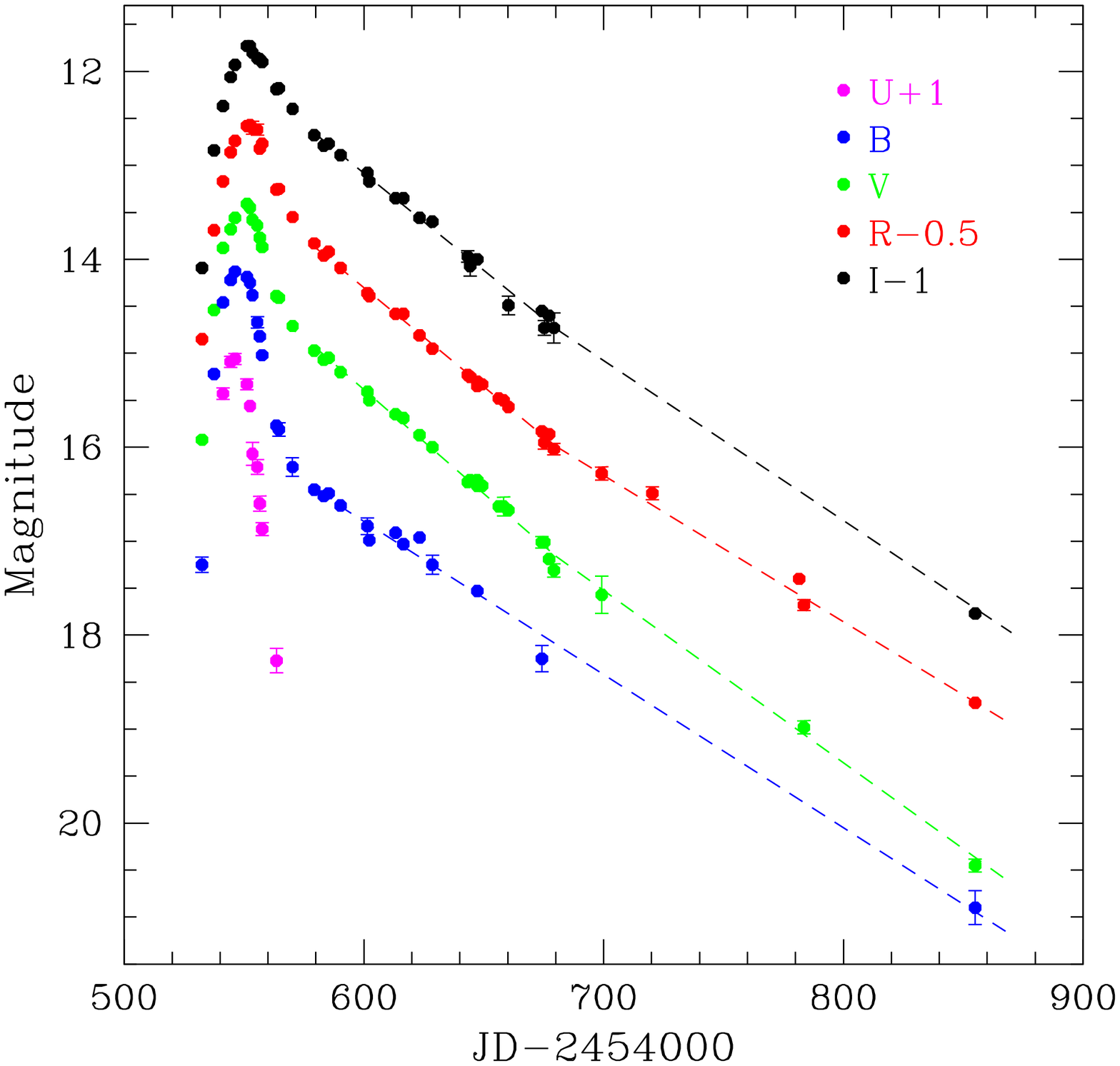}{The light curves of SN 2008ax. The dashed lines
show the decline at the tail with rates reported in the text}
The light curves of SN 2008ax are shown in Fig. 2. The premaximum 
rise and the peak have good coverage by observations, and we
can determine the dates and magnitudes of maximum light in different
bands: $U_{max}=14.06; t_{Umax}={\rm JD}2454546.9;
B_{max}=14.09; t_{Bmax}={\rm JD}2454548.2; V_{max}=13.40;
t_{Vmax}={\rm JD}2454550.3; R_{max}=13.06; t_{Rmax}={\rm JD}2454552.4;
I_{max}=12.73; t_{Imax}={\rm JD}2454552.5$.
The dates of maximum in {\it UBVRI} bands are in good agreement with
the results by P08 and R09, while for the peak
magnitudes the agreement is slightly worse.
The maximum magnitudes from R09 
in $b,v$ bands are about 0.2 mag fainter than our data, but their $u$
peak magnitude is practically equal to our estimate.
The peak $BV$ magnitudes derived by P08 are slightly fainter than our
data. 
 
After the maximum the brightness of SN declined very fast.
At the phase 15 days past maximum the $B$ magnitude declined by
1.67 mag. The fast drop continued for about 25 days, and at about 
JD 2454580 the onset of the linear decline is observed. 
The rates of decline in the period JD 2454580-680 are 
(in mag/day): 0.022 in $V$, 0.021 in $R$, 0.020 in $I$.
After JD 2454680 the rate slightly decreased, the values for the
period JD 2454680-850 are: 0.018 in $V$, 0.016 in $R$, 0.017 in $I$.
In the $B$ band the decline rate is constant for the period
JD 2454580-850 and equals 0.016. 

\PZfig{13cm}{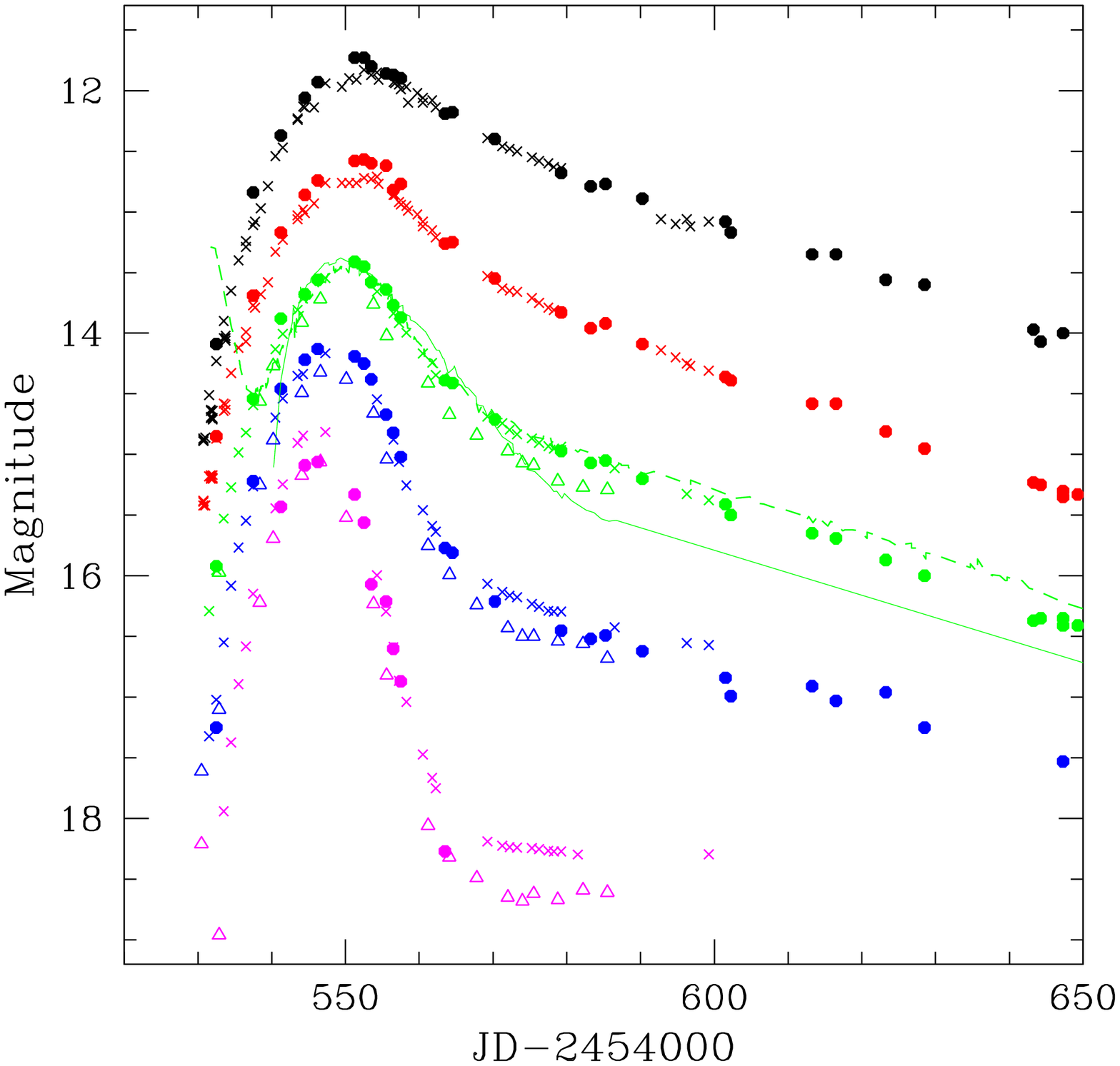}{The comparison of our data (dots) with the results
from P08 (crosses) and R09 (triangles). The color coding
and shifts are the same as on Fig. 1. Solid and dashed green lines present
the $V$-band light curves of SNe 2002ap and 1993J}
Fig. 3 presents the comparison of our data with the results by
P08 and R09. 
The photometry by P08 in Sloan
$u',r',i'$ filters is transformed to $U,R,I$ bands by the relations
derived by Chonis and Gaskell (2008). The agreement between our
magnitudes and the data from P08 is quite good, some differences are 
observed near maximum in the $U,R,I$ bands. They may result from 
errors of transformation from Sloan to Johnson-Cousins photometric
systems. The $b,v$ magnitudes from R09 are significantly
fainter than our results, while their $u$ magnitudes agree well with our $U$
data. We also plot in Fig. 3 the $V$-band light curves of SN IIb 1993J and
SN Ic 2002ap (Richmond et al, 1996; Foley et al., 2003), aligned to 
match the peaks of the curves.
The shape of the light curve of SN 2002ap is different from that of
SN 2008ax: the rise to the maximum is faster, and the decline is slower.
The second peak on the light curve of SN 1993J matches closely 
the light curve of SN 2008ax. 

The absolute $V$-magnitude light curves of SN 2008ax and several SNe
of types IIb, Ib and Ic are compared in Fig. 4.
\PZfig{13cm}{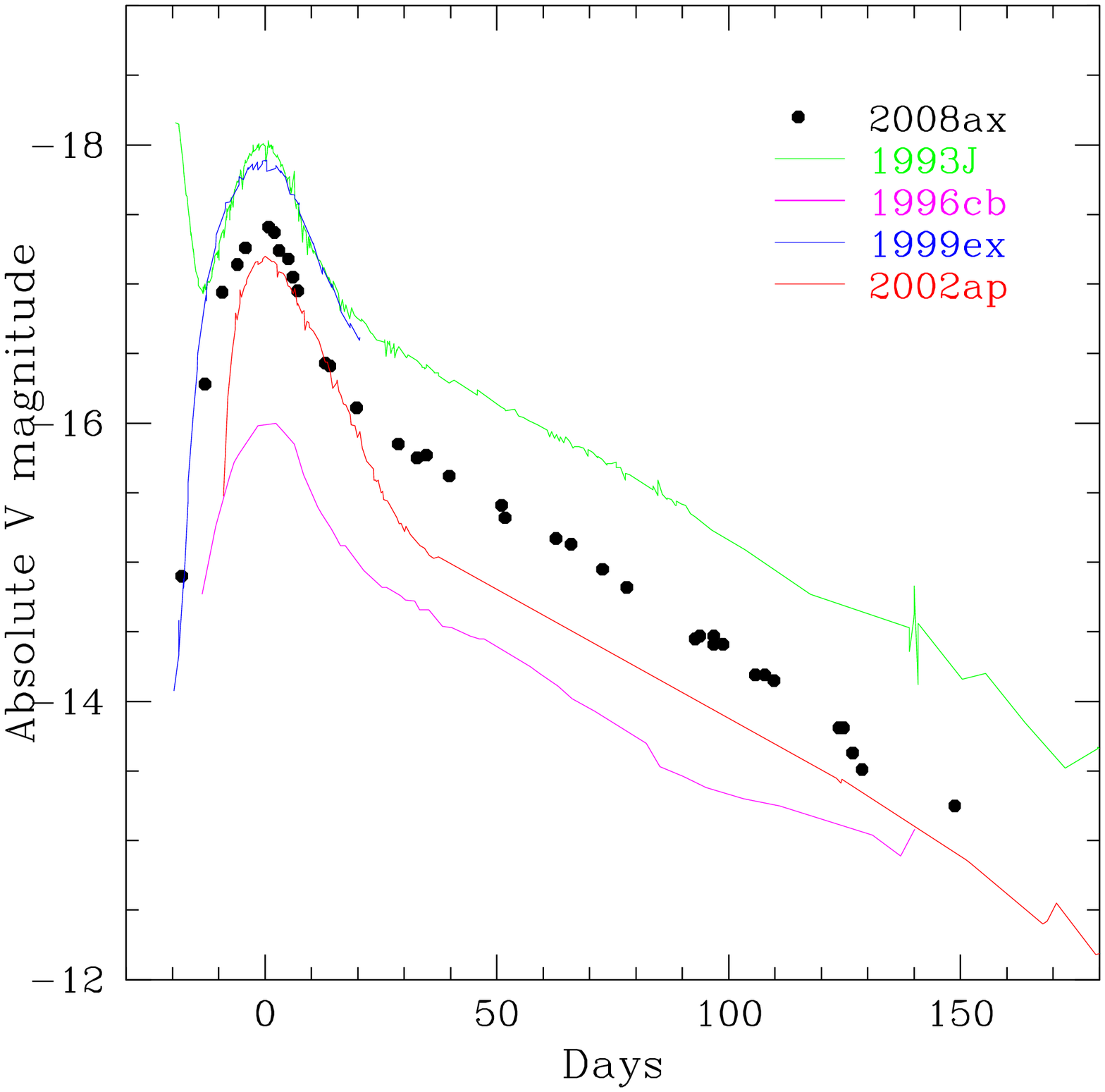}{The absolute $V$-band light
curves of SN 2008ax and SNe of types IIb (1993J, 1996cb), 
Ib (1999ex), Ic (2002ap). Day 0 corresponds to the maxima on the
light curves} 
For SN 2008ax we adopted distance modulus $\mu = 29.92$ and
extinction $E(B-V)=0.3$ as in P08. The light curves of other 
SNe are taken from Richmond et al. (1996), 
Qiu et al. (1999), Stritzinger et al. (2002), Foley et al. (2003).
With absolute peak $V$ magnitude of $-17.45$ mag SN 2008ax appears to 
be quite typical among SNe of similar classes. It is little fainter than
SN IIb 1993J and SN Ib 1999ex, have nearly the same luminosity as 
SN Ic 2002ap and 
is significantly brighter than SN IIb 1996cb.

\PZfig{13cm}{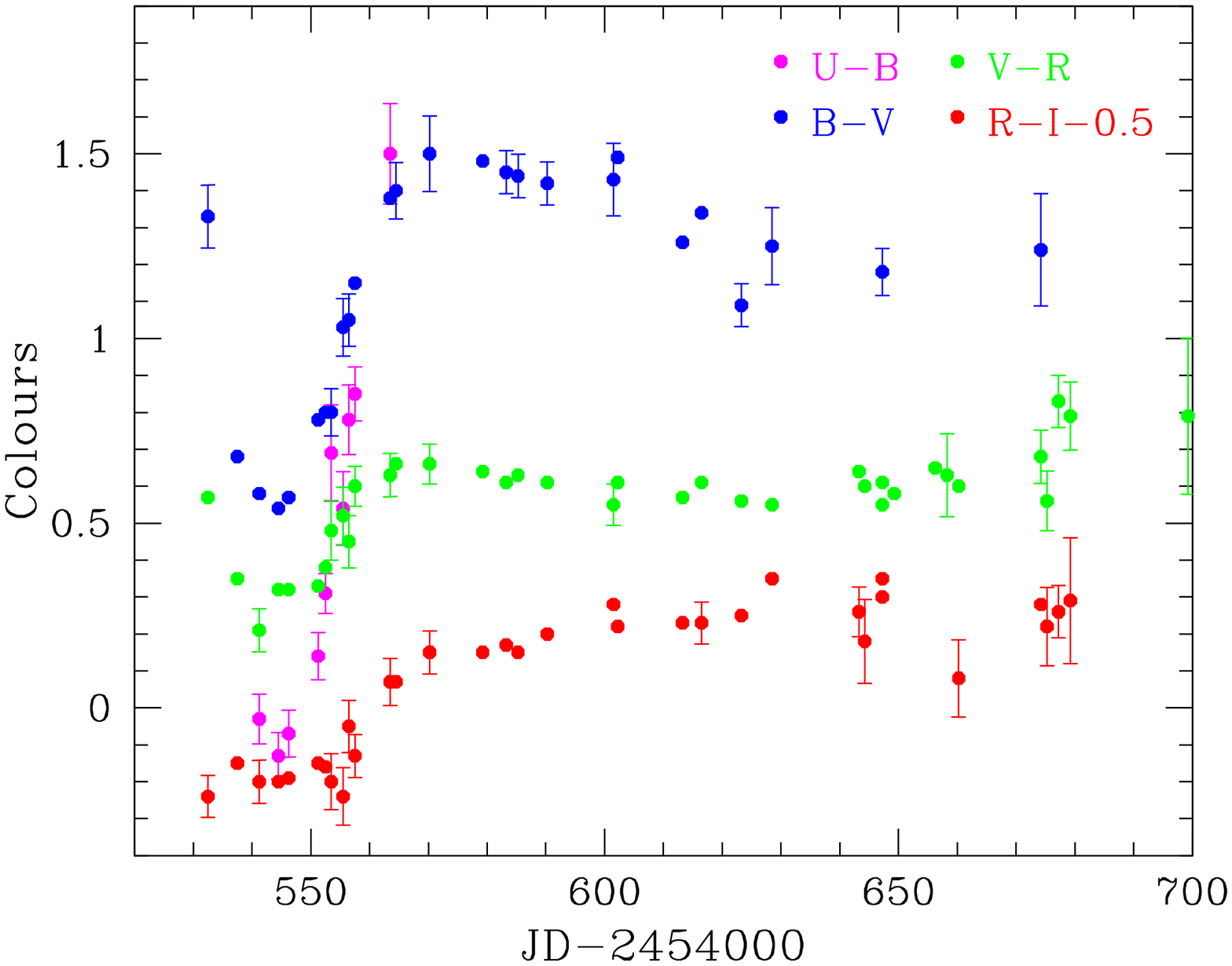}{The color curves of SN 2008ax}
 
The color curves are shown in Fig. 5. The evolution of colors
$B-V$ and $V-R$ is similar. Before maximum SN 2008ax becomes
bluer, then quickly redden, and finally the colors remain nearly
constant. $U-B$ color probably evolves in the same way, but our
data spans only the period of fast reddening. The $R-I$ color curve
is different, the color before maximum is nearly constant, and then
only a slight reddening is observed.     

\newpage
\PZsubtitle{Modeling the light curves}

We compute the light curves in {\it UBVRI} bands using
our code STELLA, which incorporates implicit hydrodynamics coupled to
a time-dependent multi-group non-equilibrium radiative transfer
(Blinnikov et al., 1998). 
The specific model employed here was Model 13C of Woosley et al. (1994). 
This model was derived from a 13 M$_{\odot}$ main sequence star that
lost most of its hydrogen envelope to a nearby companion.
The main parameters of the model are: total mass 3.8 M$_{\odot}$,
radius 600 R$_{\odot}$, mass of $^{56}$Ni 0.11 M$_{\odot}$, 
explosion energy $1.5 \times 10^{51}$ erg.
The chemical composition is shown in Fig. 6. Fig. 7 gives the resulting
light curves, and Fig. 8 presents the light curves for the first 50 days
past explosion. 

\PZfig{13cm}{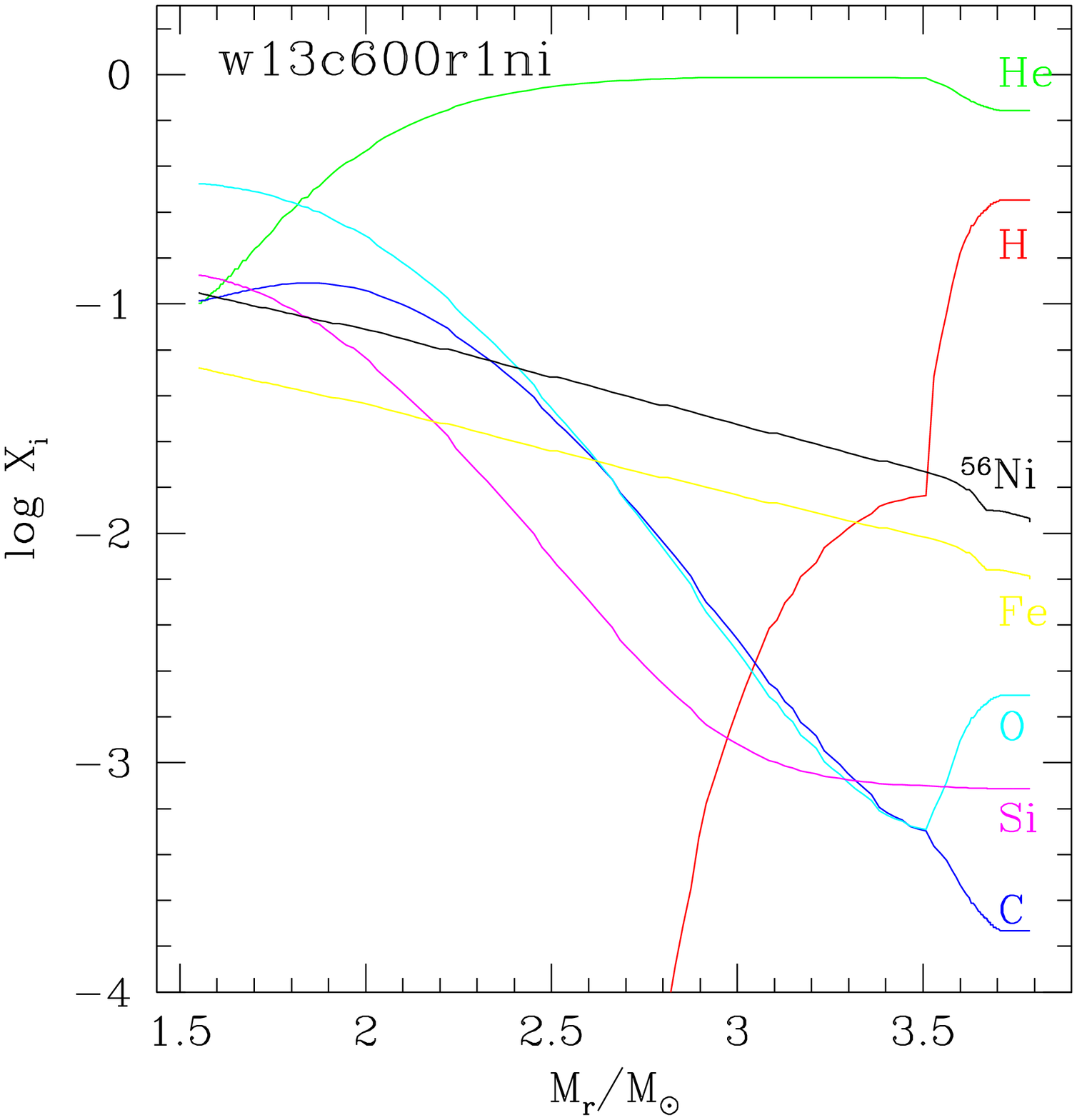}{The chemical composition of the model for light curve
computation}

\PZfig{13cm}{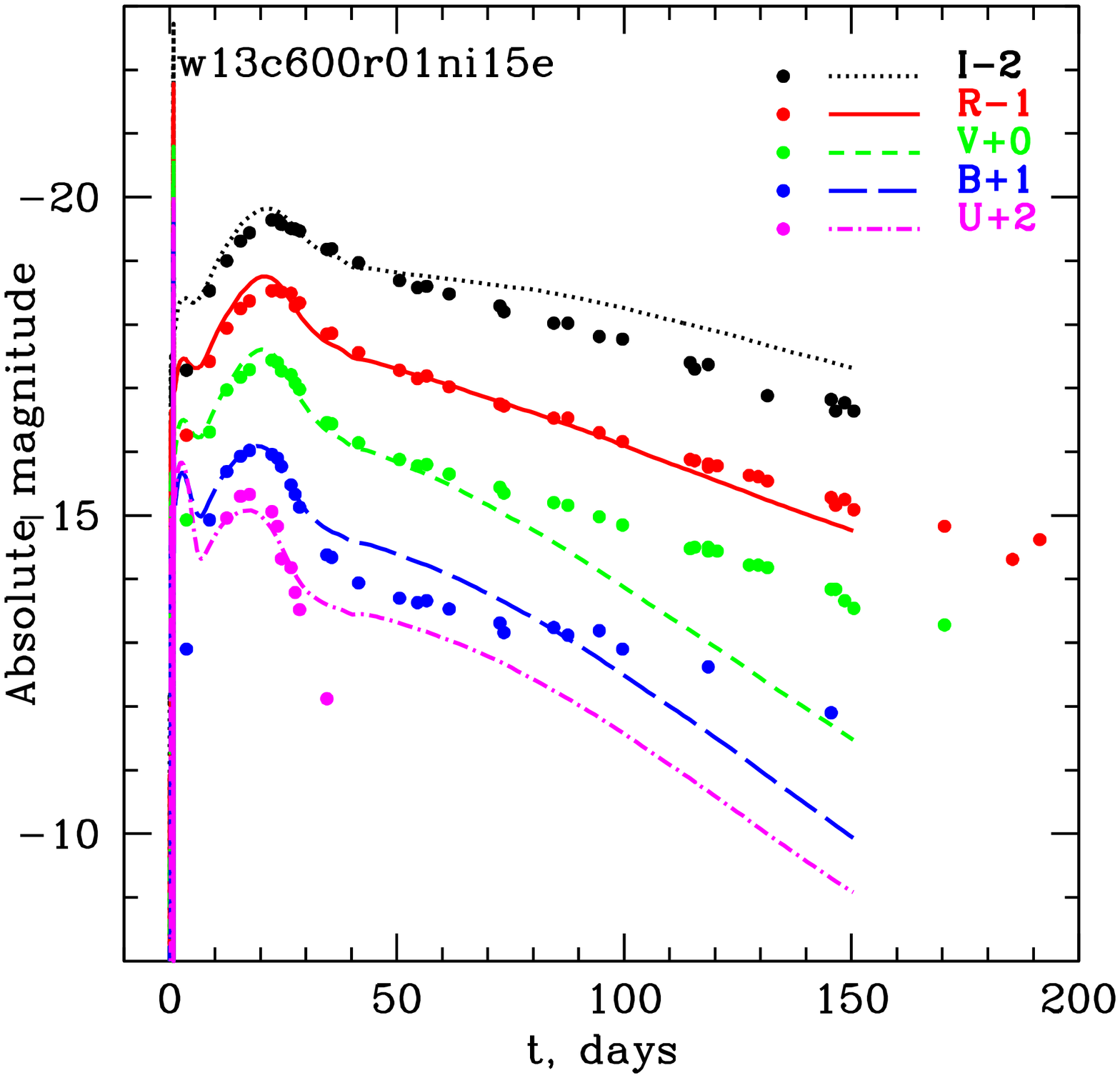}{The computed light curves compared to our observational
data. Day 0 is the time of shock breakout according to P08 (JD 2454528.8)}

\PZfig{13cm}{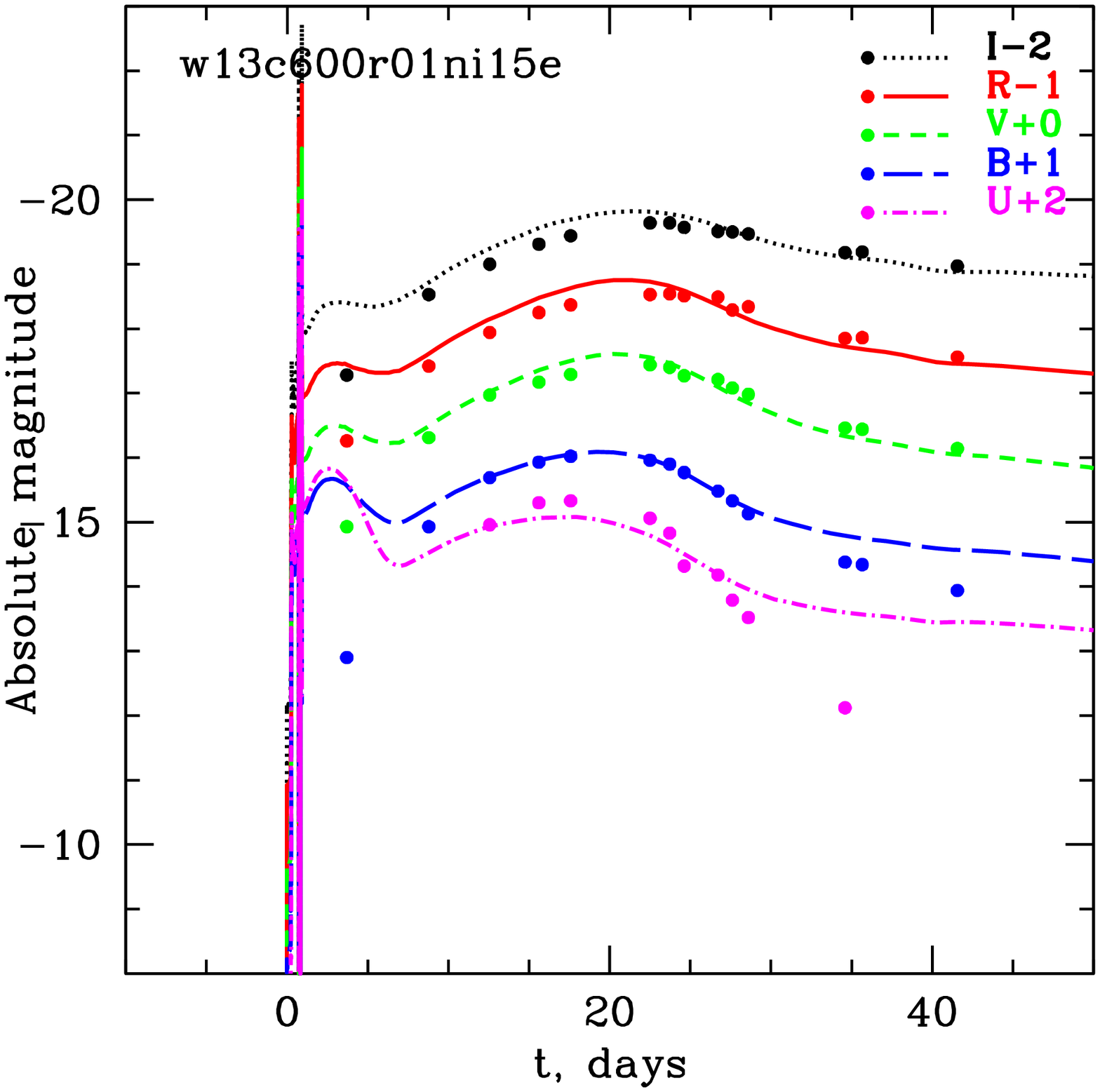}{The model light curves for the first 50 days since
explosion}

The model light curves give a very good fit to the observed maximum, 
concerning both the luminosity and the shape. The differences are on
the rising branch, where the computed early-time peak is brighter than 
observed, 
and on the tail, especially for the $U$ and $B$ bands. 
We consider the agreement to be quite satisfactory, but we continue the 
search 
for models which will give better fits. The results and more detailed
discussion of the properties of the models and their impact on the possible 
evolution of the progenitor will be published in a subsequent paper.

\bigskip
{\bf Acknowledgements.} We thank N.P.Ikonnikova and N.N.Pavlyuk, who
made some of the observations.
The work of D.T. is partly supported by the 
Leading Scientific Schools Foundation
under grant NSh.433.2008.2.
I.V. acknowledges financial support from SAI scholarship and 
from Slovak Academy Information Agency (SAIA).
The work of S.B. and P.B. is supported partly by the grant RFBR
07-02-00830-a, by the Leading Scientific Schools Foundation under
grants NSh.2977.2008.2, NSh.3884.2008.2, and in Germany by MPA guest
program. 

This paper makes use of data obtained from the Isaac Newton Group 
Archive which is maintained as part of the CASU Astronomical Data 
Centre at the Institute of Astronomy, Cambridge.

\references

Arbour, R., 2008, {\it CBET}, No. 1286

Blinnikov, S. I., Eastman, R., Bartunov, O. S., 
Popolitov, V. A., Woosley, S. E., 1998, 
{\it Astrophys. J.}, 496, 454 

Chonis, T.S., Gaskell, C.M., 2008, {\it Astron. J.}, 135, 264

Crockett, R.M., Eldridge, J.J., Smartt, S.J., et.al., 2008,
{\it MNRAS}, 391, 5 

Foley, R.J., Papenkova, M.S., Swift, B.J., Filippenko, A.V., et al.,
2003, {\it PASP}, 115, 1220

Mostardi, R., Li, W., Filippenko, A.V., 2008, {\it CBET}, No. 1280 

Nakano, S., Itagaki, K., 2008, {\it CBET}, No. 1286

Richmond, M.W., Treffers, R.R., Filippenko, A.V., Paik, Y., 1996,
{\it Astron. J.}, 112, 732

Roming, P.W.A., Pritchard, T.A., Brown, P.J., et al., 2009,
preprint (arXiv:0909.0967) 

Pastorello, A., Kasliwal, M.M., Crockett, R.M., et al., 2008,
{\it MNRAS}, 389, 955

Stritzinger, M,, Hamuy, M., Suntzeff, N.B., et al.,
2002, {\it Astron. J.}, 124, 2100

Tsvetkov, D.Yu., 1984, {\it Astron. Tsirk.}, No. 1346, 1

Woosley, S.E., Eastman, R.G., Weaver, T.A., Pinto, P.A., 1994,
{\it Astrophys. J.}, 429, 300 

Qiu, Y., Li, W., Qiao, Q., Hu, J., 1999, {\it Astron. J.}, 117, 736
 
\endreferences

\end{document}